\documentstyle{NEWmnras}
\title{\Large\bf Galaxy Redshifts: improved techniques}
\author{A.F. Heavens \\ {\small\it Department of Astronomy,
University of Edinburgh, Royal Observatory, Blackford Hill,
Edinburgh, EH9 3HJ, United Kingdom}}
\headertitle{Galaxy Redshifts: improved techniques}
\mainauthor{A.F. Heavens}
\dates{---}{---}{---}
\abstract{This paper analyses the effects of random noise in determining
errors and confidence levels for galaxy redshifts obtained by
cross-correlation techniques.  The main finding is that confidence
levels have previously been overestimated, and errors inaccurately
calculated in certain applications.
New formul\ae\ are presented.

}

\def\gs{\mathrel{\raise1.16pt\hbox{$>$}\kern-7.0pt 
\lower3.06pt\hbox{{$\scriptstyle \sim$}}}}         
\def\ls{\mathrel{\raise1.16pt\hbox{$<$}\kern-7.0pt 
\lower3.06pt\hbox{{$\scriptstyle \sim$}}}}         
\begin{document}
\maketitle

\section{INTRODUCTION}

It is becoming increasingly possible to obtain very large numbers of
galaxy spectra through the use of systems such as those involving
optical fibres.   With such large datasets, automated procedures for
obtaining redshifts become desirable and necessary for efficiency.  It
is important to  have automatic and objective measures of the confidence
levels and errors in the results, so that suspect redshifts may be
investigated individually by manual inspection of the data.   Clearly,
it is important that the initial automatic quality assessment is founded
on a firm theoretical basis.

Obtaining redshifts from galaxy spectra is commonly effected by
cross-correlating the spectrum with the spectrum of a `template' galaxy
of known redshift.  The relative shift of the two spectra when the
cross-correlation function (ccf) has its highest peak is then used to
estimate the redshift of the galaxy.  The procedure for doing this was
detailed and analysed by Tonry \& Davis (1979; hereafter TD).  In the
case of identical galaxy and template spectra the technique is
excellent, but in practice there are differences -- either
intrinsic spectral differences, or the presence of noise.  These differences
can lead to an error in the derived redshifts, or an entirely spurious
redshift being picked up.   This paper puts the analysis of noise in ccf
techniques on a firm footing, and presents improved formul\ae\ for
calculating confidence levels and errors.

\section{METHOD}

The method of obtaining redshifts from ccf techniques is presented in
TD.  One bins the galaxy spectrum $g_n$ and template $t_n$ into $N$ bins
($n=0,\ldots,N-1$) logarithmically-spaced in wavelength (initially we
follow almost the same notation as TD:  $g(n)$ in TD is written here as
$g_n$).  The spectra are continuum-subtracted and may be filtered to
remove long-wavelength components (e.g. from incomplete continuum
subtraction) and/or short wavelength contributions on
sub-resolution scales.  A cross-correlation function is then made from
the galaxy and template spectra, and the highest peak in the ccf is used
to calculate the relative shift between the two.  This is then used to
deduce the galaxy redshift, assuming the template redshift is known (see
TD for more details).

Since the ccf technique is most conveniently implemented via
Fourier methods, the spectra are operationally assumed to be periodic,
so are apodised to remove spurious short-wavelength components coming
from a mismatch at the two ends of the spectrum.

Let the discrete Fourier transforms (FTs) of $g_n$ and $t_n$ be
$G_k$ and $T_k$, where, for example
\begin{equation}
G_k = \sum_{n=0}^{N-1}\,g_n\,e^{-2\pi ink/N}\qquad k=0,\ldots,N-1.
\end{equation}
With this notation, the inverse transform is
$g_n=(1/N)\sum_k G_k e^{2\pi ink/N}$ and the variance of $g$ is
$\sigma_g^2 = (1/N)\sum_n g_n^2 = (1/N^2)\sum_k |G_k|^2$ and all sums
run from $0$ to $N-1$ unless otherwise stated.
The normalised ccf is formed:
\begin{equation}
c_n \equiv {1\over N\sigma_g\sigma_t}(g \times t)_n
\equiv {1\over N\sigma_g\sigma_t}\sum_m g_m t_{m-n}.
\end{equation}
We assume that $g_n$ is identical to $t_n$ except it is:
\begin{enumerate}
\item{different by addition of a random noise component $e_n$}
\item{multiplied by $\alpha$}
\item{shifted by $\delta$ (in bins)}
\item{broadened by convolution with a symmetric function $b_n$}
\end{enumerate}
i.e.
\begin{equation}
g_n = \alpha\left[(t * b)_{n-\delta}+ e_n\right]
\end{equation}
where the $*$ indicates convolution.  Specifically:
\begin{equation}
(t*b)_n = \sum_m\,t_m\,b_{n-m}.
\end{equation}
Transforming,
\begin{equation}
G_k=\alpha\left(T_k B_k e^{-2\pi i \delta k/N}+E_k\right)
\end{equation}
where $B_k$ and $E_k$ are the Fourier transforms of the broadening function
$b_n$ and the noise $e_n$.  The normalised ccf is
\begin{equation}
c_n =  {1\over N\sigma_g\sigma_t}\sum_m \alpha\left(\sum_p
t_p\,b_{m-\delta-p}+e_m\right)\,t_{m-n}
\end{equation}
or $c_n = \alpha\left[(t*b)\times t\right]_{n-\delta}+\left[e\times
t\right]_n$.  In Fourier space, the coefficients of $c$ are (see e.g.
Nussbaumer 1981, p.80-83):
\begin{equation}
C_k = \lambda\alpha\left(|T_k|^2B_ke^{-2\pi i\delta k/N}+T_k^* E_k\right)
\end{equation}
and $\lambda \equiv 1/(N\sigma_g\sigma_t)$.   Fig. 1 shows an example
galaxy spectrum G0102-63 and a template radial velocity standard
HD136202, with a power series continuum subtraction applied to both.
Filtering has also been applied in Fourier space to remove very
long-wavelength components from the spectra.
Also shown is the ccf, illustrating the power of the cross-correlation
technique.

We now make the principal assumption for the
following analysis: we assume the noise and the template are uncorrelated,
and $e_n$ is a random gaussian field.  Note that the noise component may
be a combination of photon noise and real spectral differences between
the two objects.   In this case,  the field specified
by $N_k\equiv T_k^* E_k$ is also a random gaussian field.   In the
absence of broadening, $\sigma_g^2=\alpha^2(\sigma_t^2+\sigma_e^2)$.

The method of estimating the noise level proposed by TD is valid for a
symmetric broadening function $b_n$.  In this case, the FT of the first
(`signal')
term in (6) is symmetric about $n=\delta$.  Hence the asymmetric part of
the actual ccf may be used to estimate the contribution from the noise.
We form the symmetric and antisymmetric parts:
\begin{equation}
s_n,a_n \equiv {1\over 2}\left(c_{\delta+n}\pm c_{\delta-n}\right)
\end{equation}
with transforms
\begin{equation}
S_k,A_k \equiv {1\over 2}\left(C_k e^{2\pi i\delta k/N}\pm
C_{-k} e^{-2\pi i\delta k/N}\right)
\end{equation}
where $C_{-k} = C_k^*$ from the reality of $c_n$.
With a symmetric $b_n$, we have $B_k^* = B_k$, so
\begin{eqnarray}
S_k & = & \lambda\alpha\left[|T_k|^2 B_k + (M_k+M_k^*)\right]\nonumber
\\
A_k & = & \lambda\alpha(M_k - M_k^*)
\end{eqnarray}
where $M_k \equiv {1\over 2}T_k^*E_k e^{2\pi i\delta k/N}$.  Thus, as stated
by TD,  the contribution to the symmetric ccf is statistically the same
as to the antisymmetric part, if the noise has random phases, uncorrelated
with the template.  The r.m.s. noise is therefore $\sqrt{2}\sigma_a$,
where $\sigma_a$ is the measured r.m.s. variation in the antisymmetric
part of the ccf.  An estimator for the FT of the broadening function may
be made by making use of this property, assuming the noise and template
are uncorrelated:
\begin{equation}
B_k^2 \simeq {|S_k|^2 - |A_k|^2 \over (\lambda\alpha)^2|T_k|^4}.
\end{equation}
The normalisation is not important here (a redefinition amounts to
altering $\alpha$).  Note that the contributions to the symmetric and
antisymmetric powers are only equal in the mean, so some averaging over
modes may be necessary in practice.


\section{CONFIDENCE ANALYSIS}

We want the probability that a peak in the ccf has been caused by noise.
It is here that we may improve substantially on the analysis of TD,
using the theory of peaks in gaussian noise originally developed by Rice
(1954).

Let the range of the ccf be $L$ bins, and let the believable fraction of
this be $f$ (one might wish, for example, to exclude negative
redshifts).  With the Fourier technique, $L=N$.  The expected number
of noise peaks, in a range $fL$,
higher than the observed peak $c_\delta$ is
\begin{equation}
\bar N = n_{pks}(>\nu) fL
\end{equation}
where $\nu = c_\delta/(\sqrt{2}\sigma_a)$ is the height of the ccf peak
in units of the r.m.s. noise.  $n_{pks}(>\nu)$ is the number density of
peaks above $\nu$, and is given by (Williams {\it et al.} 1991):
\begin{eqnarray}
n_{pks}(>\nu) = {1\over 4\pi R}\left\{{\rm erfc}\left[{\nu\over
\sqrt{2(1-\gamma^2)}}\right]+\right.\nonumber \\
\left.\gamma e^{-\nu^2/2}\left[1+{\rm erf}\left(
{\gamma\nu\over\sqrt{2(1-\gamma^2)}}\right)\right]\right\}.
\end{eqnarray}
Note that $\nu$ here is identical to $r$ in TD. The quantities
$R \equiv \sigma_1/\sigma_2$ and $\gamma \equiv
\sigma_1^2/(\sigma_0\sigma_2)$ characterise the noise in terms of
moments of its power spectrum:
\begin{equation}
\sigma_j^2 \equiv {2\over N^2}\sum_k^{N/2} |N_k|^2 \left({2\pi k\over
N}\right)^{2j}.
\end{equation}
Note that the sum extends over half the space.  Reality of the noise
ensures each half contributes equally.  $R$ is a length which
characterises the coherence properties of the noise, and $\gamma$ is a
dimensionless parameter which measures the relative contribution from
short and long wavelengths.

To obtain the confidence level $C$ for the observed peak in the ccf, we need
the probability that there are no higher noise peaks in the interval.
This is a non-trivial task, and the theory of peaks cannot currently
provide an answer.  However, a good approximation for high peaks
is to assume that the peaks are uncorrelated (Williams {\it et al.}
1991), in which case
\begin{equation}
C \simeq e^{-{\bar N}} = e^{-n_{pks}(>\nu)fL}.
\end{equation}
This should be very accurate if $\bar N \ll 1$.  If not, the confidence
is low, and the redshift obtained will be suspect anyway.

To obtain the confidence level, we need to estimate the parameters $R$
and $\gamma$ from the noise.  This may be done either by performing an
FT on the antisymmetric part of the ccf, and using (14), or by making
use of the following results, which follow from (13):
\begin{equation}
n_{pks}(>-\infty) = {1\over 2\pi R};\qquad n_{pks}(>0) = {1+\gamma\over
4\pi R}.
\end{equation}
The total number of peaks $N_{pks}$, and the number of positive peaks
$N_{pks}(>0)$ may therefore be
used more straightforwardly to estimate $R$ and $\gamma$:
\begin{equation}
\gamma \simeq {2\,N_{pks}(>0)\over N_{pks}}-1;\qquad
R \simeq {L\over 2\pi N_{pks}}.
\end{equation}
The two methods give, in tests involving nearly a thousand ccfs, the
same answers to an r.m.s. accuracy of  3\% for $R$.  The accuracy in
$\gamma$ is not so high,  7\% to 24\% depending on whether the spectra
are filtered or not.  The Fourier method is to be preferred.

For high peaks ($\nu \gg 1$), one may make use of the asymptotic
expansion ${\rm erfc(z)} \rightarrow e^{-z^2}/(\sqrt{\pi}z)$ to get
an answer good to 3\% for $\nu > 2$ and $\gamma>0.5$:
\begin{equation}
C \simeq 1-2f\left[N_{pks}(>0) - {1\over 2}N_{pks}\right]e^{-\nu^2/2}
\quad (\nu\gg 1).
\end{equation}
This may be compared with TD's high-peak limit $C_{TD} = 1-\sqrt{2/\pi}f
N_{pks}(>0) e^{-\nu^2/2}/\nu$.  TD's analysis thus overestimates
the confidence by an amount
\begin{equation}
{1-C\over 1-C_{TD}}= \sqrt{2\pi}{\nu\gamma\over 1+\gamma}\qquad (\nu\gg
1).
\end{equation}
These confidence estimates have been compared with TD's for a sample of
around 80 galaxies with redshifts $\ls 0.05$, each correlated with 8
radial velocity standard stars and 2 nearby galaxies.  The
signal-to-noise of the ccfs is generally high, with only one with $\nu <
2.5$ and most with $\nu > 5$.   For this test sample, the Tonry and
Davis method systematically overestimates the confidence level
(see Fig. 2).   As an indicator, a 95\%\ confidence level under the
previous analysis corresponds roughly to a true confidence of 84\%.

\section{ERRORS}

If the correct peak has been selected, the redshift may still have an
error as the noise can move the position of the maximum.   The function
\begin{equation}
c_n  = {\lambda\alpha\over N}\sum \left(|T_k|^2B_k e^{-2\pi
ik\delta/N}+T_k^*E_k\right)e^{2\pi ikn/N}
\end{equation}
is maximised.  The first term ($\equiv \hat c$) is symmetric, peaking at
$n=\delta$.  Expanding this term in a Taylor expansion, and writing
the second term as $\epsilon$, the peak is shifted to
$n-\delta = -\epsilon'/\hat c_\delta''$, the primes indicating
derivatives.  Thus, for an ensemble of ccfs with the same template but
different noise realisations,  the maxima are shifted
by an r.m.s. amount
\begin{equation}
\Delta x = \sqrt{2}\sigma_{1a}/(-\hat c_\delta'').
\end{equation}
where we have used the antisymmetric part of the ccf to write
$\sigma^2_{1\epsilon} = 2\sigma^2_{1a}$.
The second derivative at the peak may be written
\begin{equation}
-\hat c_\delta'' = \lambda\alpha{(2\pi)^2\over
N^3}2\sum_{k=1}^{N/2}k^2\,|T_k|^2B_k
\end{equation}
To make further progress requires specific assumptions about the
broadening function.  Broadening by convolution with a gaussian
is considered in Section 5, which demonstrates that the error is rather
insensitive to broadening.  Here therefore I consider
no broadening ($B_k=1$), in which case,
\begin{equation}
-\hat c_\delta'' = \lambda\alpha N \sigma_{1t}^2 = {\alpha\sigma_{1t}^2\over
\sigma_g\sigma_t}
\end{equation}
where $\sigma_{1t}^2$ is the second moment of the filtered template
spectrum.  Calculation of $\sigma_g$ requires $\sigma_e$, which may be
calculated by noting that $\lambda\alpha|E_k| = \sqrt{2}|A_k|/|T_k|$,
from which we get
\begin{equation}
\sigma_e^2+\sigma_t^2 = {\sigma_t^2\over 1-2\sigma_t^2\sum\left(
|A_k|^2/|T_k|^2\right)}
\end{equation}
So we obtain the final expression for the r.m.s. error:
\begin{equation}
\Delta x = \sqrt{2}{\sigma_{1a}\sigma_t^2\over
\sigma_{1t}^2}\left[1-2\sigma_t^2\sum\left(|A_k|^2/|T_k|^2\right)
\right]^{-1/2}.
\end{equation}
This can be tested in two ways.  The first test is to do exactly what
this analysis assumes:  a template spectrum has noise added, and is
shifted.  The ccf technique is then used to estimate the shift.  Fig. 3a
shows the distribution of errors, normalised to the error estimate (25),
along with the expected gaussian of unit variance, for white
noise. Slight deviations are expected due to the
discrete sampling of the spectra, but the agreement is remarkably good.
Differences may arise because,  although the peak of the ccf is
calculated to sub-pixel accuracy, the antisymmetric part is calculated
assuming $\delta$ is an integer.

A further weak test of the method is shown in Fig. 3b, which is based on
real data;  if the differences between galaxy spectra and templates  can
be characterised in the way suggested in this paper, the errors  should
be distributed normally.  The figure shows the distribution of  errors,
each normalised to the r.m.s. (for each galaxy  cross-correlated to the
ten templates), and the ideal case of a gaussian of unit variance.
Cases where the wrong ccf peak has been selected are
almost always easy to spot and these have been removed.
Note that it is incorrect to use the standard error for the mean of the
$n$ redshifts (r.m.s./$\sqrt{n}$ if all errors are the same), as the
noise in the galaxy spectrum may shift the derived redshift
systematically for all templates.

\subsection{Filtering}

The results shown so far have used a FIGARO routine SCROSS, which has
been generalised to include errors (available on some STARLINK sites as
XCORR).  SCROSS filters out low-frequency components after continuum
subtraction, whereas XCORR offers the user the option not to do this. In
XCORR, the errors are estimated from the actual full-width at
half-maximum  $W$ of the peak in the ccf: error $= 0.2833 W/(1+\nu)$. A
similar  routine, XCSAO, written for IRAF (Kurtz et al. 1992) uses the
formula $0.375W/(1+\nu)$.  These work reasonably well in practice, but
there can be cases where noise broadens the wings of the peak
substantially, while the peak remains locally sharp, as illustrated in
Fig. 4.  The error in such cases can be grossly overestimated.  Fig. 1
shows the filtered case, for the same spectra. Here the actual width is
a better guide, but sometimes provides a poor error estimate, usually an
overestimate.  The formula (25) is based on local properties at the peak
itself, so is insensitive to any broad wings.

\section{Gaussian Power Spectra}

In the ideal case of no noise and no broadening, the normalised ccf has
a peak of 1.   In practice, the height of the peak falls below this, and
one ought to be able to use the actual height to estimate the
significance of the redshift obtained.   This is indeed the case, and
one can get a rule-of-thumb estimate by making a specific assumption
that the power spectra involved are gaussian.  If the spectra are
filtered, this approximation is usually quite good.

We assume the broadening function is a gaussian, with width $\sigma$,
that the template and noise have gaussian power spectra with
characteristic widths $\tau$ and $\eta$. i.e.
\begin{eqnarray}
B_k & = & \exp\left(-{2\pi^2 k^2 \sigma^2\over N^2}\right)\\
|T_k|^2 & = & 2\sqrt{\pi} N \tau \sigma_t^2 \exp\left[-\left(
{2\pi k \tau\over N}\right)^2\right]\\
|E_k|^2 & = & 2\sqrt{\pi} N \eta \sigma_e^2 \exp\left[-\left(
{2\pi k \eta\over N}\right)^2\right].
\end{eqnarray}
A result useful for the estimators below is
\begin{eqnarray}
2 \sum_0^{N/2}k^n \exp(-\beta k^2) & \simeq & \int_{-\infty}^\infty k^n
\exp(-\beta k^2) dk\\
& = & \Gamma\left({n+1\over 2}\right) \beta^{-(n+1)/2}.
\end{eqnarray}
Evaluating the sums in sections 2 and 3, we find
\begin{equation}
\sigma_g^2 = \alpha^2\left(\sigma_t^2{\tau\over\sqrt{\tau^2+\sigma^2}} +
\sigma_e^2\right).
\end{equation}
%
%
The mean value of the second derivative obeys
\begin{equation}
<-c''_\delta> = {\sqrt{2}\tau\sigma_t\over
\left(2\tau^2+\sigma^2\right)^{3/2}
\sqrt{\sigma_t^2{\tau\over \sqrt{\tau^2+\sigma^2}}+\sigma_e^2}}.
\end{equation}
and if we approximate the ccf peak as a gaussian, $c_n \simeq
c(\delta)\exp\left[-(n-\delta)^2/(2\mu^2)\right]$,  matching second
derivatives at the peak gives
\begin{equation}
\sigma^2 = \mu^2 - 2\tau^2
\end{equation}
in agreement with the maximum likelihood method of TD.  From this, one
can obtain the velocity broadening in the galaxy provided any
additional filtering is accounted for.

For gaussian power spectra, the error may be written
\begin{equation}
\Delta x = {2\pi^{1/4}\eta\tau^2\over N^{1/2}(\tau^2+\eta^2)^{3/4}}
{\sigma_e\over\sigma_t}.
\end{equation}
To evaluate this, we note that the noise term in the ccf has a variance
\begin{equation}
\sigma_\epsilon^2 = {2\sqrt{\pi}\,\tau\eta\, \sigma_e^2\over
N\left(\sigma_t^2{\tau\over \sqrt{\tau^2+\sigma^2}}+\sigma_e^2\right)
\sqrt{\tau^2+\eta^2}}.
\end{equation}
and we may use this to eliminate $\sigma_e/\sigma_t$ from (25).  After
some algebra, (25) becomes
\begin{eqnarray}
\Delta x & = & {2 \pi^{1/4}\eta\tau^{5/2} \over N^{1/2}
(\tau^2+\eta^2)^{3/4}(\tau^2+\sigma^2)^{1/4}}\nonumber \\
& \times & \left[{2\pi^{1/2}\tau\eta\over
N\sigma_{\epsilon}^2(\tau^2+\eta^2)^{1/2}}-1\right]^{-1/2}.
\end{eqnarray}
For high peaks this reduces to
\begin{equation}
\Delta x = \left({2\eta\tau\over \tau^2+\eta^2}\right)^{1/2}
\left[{\tau^2\over \tau^2+\sigma^2}\right]^{1/4}\tau\sigma_{\epsilon}
\end{equation}
which demonstrates the insensitivity of the error to the broadening.
One can write the error in terms of the number of peaks, with the hope
that the result can be used generally.   If one assumes $\eta \simeq
\tau \gg \sigma$ and uses
\begin{equation}
N_{pks} = {\sqrt{3}\over 2^{3/2} \pi}{N\over \sqrt{\tau^2+\eta^2}}.
\end{equation}
then the error may be written
\begin{equation}
\Delta x \simeq {\sqrt{3}N\sigma_\epsilon\over 4\pi N_{pks}}
\end{equation}
Unforturtunately, in our sample, this turns out to be a poor estimator,
underestimating errors by a factor of about 3.

\section{CONCLUSIONS}

This paper has analysed the effects of noise on the confidence levels
and errors in the cross-correlation technique widely used to obtain
galaxy redshifts.  This improved treatment finds somewhat lower
confidence levels than previously found, and also provides a formally
more correct error assessment for the redshifts obtained.    The
principal results are the confidence level (equation 15 with 13, or the
approximation, equation 18), and the error estimate (equation 25).

The results should be particularly useful in identifying questionable
redshifts in programmes which obtain large numbers of galaxy spectra.
In particular, a low confidence level would indicate that the object
should be checked carefully before the redshift is accepted. It should
be borne in mind that errors may well arise from factors which are not
treated in this paper, principally wavelength calibration errors  and
spectral differences between galaxies and templates which cannot  be
described by noise (even when filtered).

It is hoped to provide a FIGARO version of the cross-correlation
program SCROSS to provide error and confidence analysis.

\vspace{0.5cm}

\noindent{\bf Acknowledgments}

\noindent I am grateful to Andy Connolly for providing galaxy spectra,
and for useful comments on the manuscript, to an anonymous referee for
some helpful remarks,  and to John Peacock for a useful point.  Computer
processing was done using STARLINK facilities.

\vspace{0.5cm}

\noindent{\bf References}

\begin{description}
\item
Kurtz M.J., Mink D.J., Wyatt W.F., Fabricant D.G., Torres G., Kriss
G.A., Tonry J.L., 1992, in Worrall D.M., Biemesderfer C., Barnes J.,
eds,  Astronomical Data Analysis
\& Systems I, (A.S.P. Conference Series vol. 25), p.432
\item
Nussbaumer H.J., 1981 Fast Fourier Transforms and Convolution
Algorithms.  Springer-Verlag, Berlin
\item
Rice S.O., 1954, in Wax N., ed, Selected Papers on Noise and
Stochastic Processes, p.133. Dover, New York
\item
Tonry J., Davis M., 1979, AJ, 84, 1511
\item
Williams B.G, Heavens A.F., Peacock J.A., Shandarin S.F., 1991,
MNRAS, 250, 458
\end{description}

\noindent{\bf Figure Captions}

\noindent{\bf Figure 1.} Sample spectra of a galaxy and template star, and
their
cross-correlation function.  In the cross-correlation process, a Fourier
filter has been applied to remove long-wavelength components.

\noindent{\bf Figure 2.}   Confidence assessments compared with the method of
Tonry and Davis for about 800 filtered ccfs, based on around
80 galaxies with redshifts $\ls 0.05$, and 10 templates, 8 of which are
radial velocity standard stars and the other two nearby galaxies.  The
spectra are mostly good signal-to-noise, with all ccf peak heights
except one above $\nu = 2.5$, and about 2/3 above $\nu = 5$.

\noindent{\bf Figure 3.}  The distribution of errors, normalised to the
sample standard deviation by a) adding white noise to a template and
cross-correlating in the normal way.  The error plotted is divided by
the program error estimate (25) and the distribution should follow the
solid curve b) cross-correlating real galaxies  with 10 templates.  The
error estimate in this case is the r.m.s. deviation from the mean of the
templates. If the assumptions in the analysis were strictly realised in
practice, the distribution should be normal (solid curve).

\noindent{\bf Figure 4.} The ccf for the galaxy-template pair of figure 1, but
with no long-wavelength filtering applied.  The actual FWHM of the peak is a
poor estimate of the error.

\end{document}